\documentclass[prl,twocolumn,showpacs,superscriptaddress,floatfix]{revtex4}
\usepackage{graphicx}

\begin{document}

\title{Dynamics of Localized Waves}

\author{Z.Q.~Zhang}
\affiliation{Department of Physics, Hong Kong University of Science and Technology,
Clear Water Bay, Kowloon, Hong Kong}

\author{A.A.~Chabanov}
\affiliation{Department of Physics and Astronomy,
University of Texas, San Antonio, Texas 78249, USA}

\author{S.K.~Cheung}
\affiliation{Department of Physics, Hong Kong University of Science and Technology,
Clear Water Bay, Kowloon, Hong Kong}

\author{C.H.~Wong}
\affiliation{Department of Physics, Hong Kong University of Science and Technology,
Clear Water Bay, Kowloon, Hong Kong}

\author{A.Z.~Genack}
\affiliation{Department of Physics, Queens College of the City University
of New York, Flushing, New York 11367, USA}

\date{\today}

\begin{abstract}
We have measured pulsed microwave transmission through quasi-1D samples with lengths up to three localization lengths. For times approaching four times the diffusion time $\tau_D$, transmission is diffusive in accord with the self-consistent theory of localization for the renormalized diffusion coefficient in space and frequency, $D(z,\Omega)$. For longer times, the transmission decay rate first agrees with and later falls increasingly below the self-consistent theory. Beyond the Heisenberg time, the decay rate approaches the predictions of a dynamic single parameter scaling model which reflects the decay of long-lived localized modes and converges to the results of 1D simulations.
\end{abstract}


\maketitle
The theory of localization was developed in the context of electronic conduction and has been widely applied to steady state transport \cite{Anderson,Mesobook}. Because of electron-electron interactions, however, the localization transition is not a pure single-particle Anderson transition. The description of localization is made all the more complex by inelastic scattering, since the impact of scattering grows with pathlength and waves following paths of all lengths contribute to the conductance. Localization can also be studied for classical waves in optics and acoustics \cite{Azbel}. However, the exponential decay of transmission found in the presence of absorption even for diffusive waves makes it difficult to determine the localization length from measurements of the exponential scaling of transmission \cite{Azi,Wiersma_N,Maret_N}. Similarly, the localization length cannot be determined directly from the rounding of the coherent backscattering peak which is produced by both absorption and localization \cite{Wiersma_N,Maret_N}.

The ensemble average of pulsed transmission, $\langle I(t)\rangle$, is of particular interest for classical waves \cite{Diff_Dynamics} because it allows the increasing impact of weak localization to be disentangled from absorption, since the relative weights of paths within the sample at a given delay time is unaffected by absorption \cite{Weaver,Nature,Chabanov1,Maret}. Far from the localization threshold, measurements of $\langle I(t)\rangle$ in opaque samples have generally been well described by diffusion theory \cite{Diff_Dynamics}. For times greater than the diffusion time, $\tau_D$, higher diffusion modes decay rapidly leaving energy in the lowest diffusion mode so that the decay rate approaches the constant, $1/\tau_D = \pi^2D/(L + 2z_0)^2$ \cite{Diff_Dynamics}. Here, $D$ is the diffusion coefficient, $L$ is the sample length, and $z_0$ is the distance beyond the boundary at which the intensity within the sample extrapolates to zero \cite{Zhu}. However, a progressive suppression of the decay rate has been observed in recent microwave \cite{Chabanov1} and optical \cite{Maret} measurements in strongly scattering samples in which steady state transmission is essentially diffusive.

The impact of localization on electron dynamics for single electrons at $T$=0 had been calculated using diagrammatic, nonlinear $\sigma$, and supersymmetry approaches \cite{Altshuler}. Localization is achieved when the average spacing between quasimodes exceeds their average linewidth, $\Delta\nu$$>$$\delta\nu$ \cite{Thouless}. In the time domain, this is the condition that the Thouless time exceeds the Heisenberg time, $\tau_{\rm Th}$$>$$\tau_{\rm H}$, where $\tau_{\rm Th}$=$1/\delta\nu$=$\pi^2\tau_D$, and $\tau_{\rm H}$=$1/\Delta\nu$ is the time required to visit each coherence volume of the sample.

The slowing decay of $\langle I(t)\rangle$ reflects the increasing prominence of longer-lived modes which are more remote from the sample boundaries or are more sharply peaked within the sample \cite{Azbel,Pendry,Sebbah,Wiersma}. This is associated with the increasing enhancement of weak localization with longer pathlength due to the scattering of the wave as it crosses over its trajectory. Vollhardt and Wolfle (VW) developed a self-consistent diagrammatic theory of localization within a medium in terms of a frequency-dependent renormalized diffusion coefficient, $D(\Omega)$ \cite{SC}. To be self-consistent, Van Tiggelen {\it et al.} \cite{Bart} argued that $D$ must also be a function of depth within a bounded sample, $D(z,\Omega)$. Skipetrov and Van Tiggelen used the self-consistent localization theory (SCLT) to describe waves near the mobility edge for $t$$<$$\tau_{\rm H}$ in quasi-1D \cite{Bart1} and slab geometries \cite{Bart2}. They described \cite{Bart1} key features observed in microwave measurements for diffusive waves in quasi-1D \cite{Chabanov1} and found a $1/t^2$ falloff in reflection for localized waves.

In this Letter, we present microwave measurements of dynamic transmission for localized waves in quasi-1D samples with $L$ greater than the average localization length, $\bar{\xi}$. Four different approaches have been used to analyze the measurements in different time ranges. For times up to several times the peak arrival times, $t_p$$\sim$$\tau_D$, $\langle I(t)\rangle$ can be well described by a simple diffusion theory. These results are in accord with the SCLT, which includes a position and frequency dependent diffusion coefficient $D(z,\Omega)$ \cite{Bart1,Bart2}, and suggest that the renormalization of $D$ is insignificant at early times. For $t$$>$$4\tau_D$, the transmission decay rate, $\Lambda(t)$=$-(d\langle I(t)\rangle/dt)/\langle I(t)\rangle$=$-d\ln\langle I(t)\rangle/dt$, is progressively suppressed by localization up to a factor of nearly 3 relative to the early diffusive decay rate, $1/\tau_D$.  Measurements of $\Lambda(t)$ are compared to the SCLT, to 1D simulations, and to a dynamic single parameter scaling (SPS) model. This model is based upon a Gaussian distribution of Lyapunov exponents, $\gamma$=1/2$\xi$, with ${\rm var}(\gamma)$=$\bar{\gamma}/L$ \cite{SPS}. Self-consistent calculations provide reasonable agreement with measurements for $t$$<$$4\tau_D$, but give substantially higher values for $\Lambda(t)$ at longer times. 1D simulations give a peak in the decay rate which is higher and peaks later than measurements. The dynamic SPS model rises within the pulse width and falls below measurements for $t$$<$$\tau_{\rm H}$. The diffusion-like delay of the transmission peak reflects the impact of short-lived, spectrally-overlapping, quasi-extended quasimodes, described by Pendry as necklace states \cite{Pendry}. These results indicate the greater prominence of overlapping modes in 1D than in quasi-1D samples. For $t$$>$$\tau_{\rm H}$, the results of 1D simulations and the dynamic SPS model which reflect the contributions of long-lived modes with $\xi$$<$$\bar{\xi}$ converge and are in good agreement with measurements.

Microwave spectra of the field transmitted through low-density random mixtures of alumina spheres were taken with the use of a vector network analyzer. The wave is launched and detected by conical horns placed 30 cm in front of and behind the sample. Alumina spheres with diameter 0.95 cm and index of refraction 3.14 are embedded within Styrofoam shells to produce a sample with alumina volume fraction 0.068 which displays distinct sphere resonances \cite{PRL2001}. The sample is contained within a copper tube with diameter of 7.3 cm and plastic end pieces. Spectra are taken for 10,000 configurations produced by briefly rotating the tube, in samples of length 13, 29, 40, 50, 61, and 90 cm. Measurements are made just above the first sphere resonance over the frequency range 9.95-10.15 GHz, in which the change in static and dynamic propagation parameters is small.

The time response to a Gaussian intensity pulse of width $(2\sqrt{2}\pi\sigma)^{-1}$ peaked at $t$=0 was obtained by taking the Fourier transform of the field spectrum multiplied by a Gaussian envelope of width $\sigma$ centered at $f_c$. The field of the time response is squared to give the transmitted intensity $I(t)$ for each sample realization. The average transmitted intensity $\langle I(t)\rangle$ is found by averaging over the ensemble, and then over the frequency interval by shifting $f_c$. The measured time response includes a constant background at long times \cite{Chabanov1}, which results from noise in the field spectra. Subtracting this background enhanced the dynamic range by 16 dB.

To compensate for losses due to absorption, and thus to facilitate the comparison of the measurements to dynamical models of localization, $\langle I(t)\rangle$ is multiplied by $\exp(t/\tau_a)$, where $1/\tau_a$ is the absorption rate. $1/\tau_a$=0.0064 ns$^{-1}$ is found from the decay rate of transmission in a 40cm-long sample with copper end caps, which is weakly coupled to the measurement ports, so that the leakage rate is well below the absorption rate [Fig.~1a].
\begin{figure}[t!]
\includegraphics[width=\columnwidth]{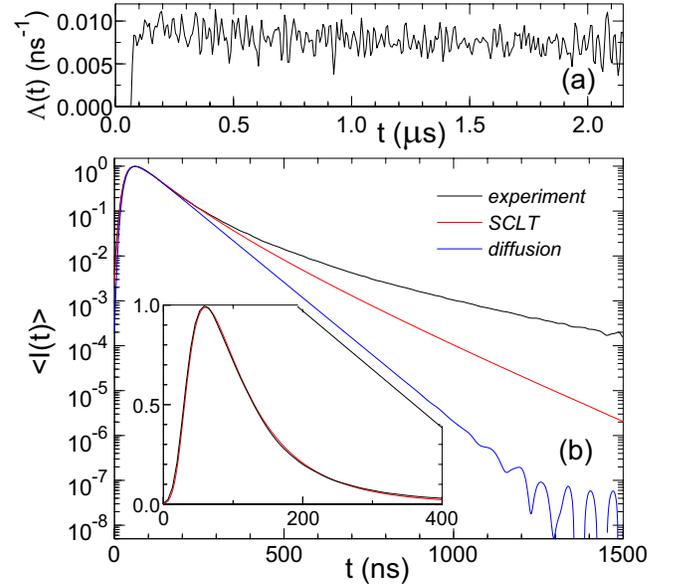}
\caption{(color on line) (a) The transmission decay rate in a $L$=40 cm sample weakly coupled to the measurement ports, so that the leakage rate is well below the absorption rate $1/\tau_a$. $1/\tau_a$=$0.0064\,$ns$^{-1}$ is found as the average of $\Lambda(t)$ for $t$$>$1.2 $\mu$s, multiplied by 0.91 to correct for the copper end caps added to the tube; (b) The average time response to a Gaussian pulse of $\sigma$=15 MHz in the $L$=61 cm sample is compared to the SCLT and diffusion theory. All the curves are normalized to be unity at the peak. The inset shows a fit of the SCLT to the measured data at early times.}
\end{figure}

Measurements are compared to calculations of the renormalized diffusion constant, $D(z,\Omega)$. In an open system, the SCLT of VW \cite{SC} can be generalized as follows \cite{Zhang},
\begin{equation}
{1\over D(z,\Omega)}={1\over D_B}\left[1+{v_E\over 2N}\, G(z,z;\Omega)\right],
\label{}
\end{equation}
where $D_B$=$v_E\ell/3$ is the Boltzmann diffusion constant, $\ell$ is the transport mean free path, $N$=32 is the number of transverse propagating channels in the sample, and $v_E$=11.85 cm/ns is the transport velocity at 10 GHz \cite{Ad}. The diagonal intensity Green function, $G(z,z;\Omega)$, represents the return probability at $z$ and can be obtained from the following generalized diffusion equation,
\begin{equation}
\partial_z[D(z,\Omega)\, \partial_zG(z,z';\Omega)]+i\Omega\, G(z,z';\Omega)=-\delta(z-z'),
\label{}
\end{equation}
with mixed boundary conditions at both ends, $z_0D(z_b;\Omega)\partial_zG(z_b,z';\Omega)\mp D_BG(z_b,z';\Omega)$=0, where $z_b$=0 or $L$, $z_0$=$(2/3)\ell(1$$+$$R)/(1$$-$$R)$ is the extrapolation length, and $R$ is the internal reflection coefficient. Equations~(1) and (2) are solved self-consistently in real $\Omega$-space to obtain $G(z,z';\Omega)$. The intensity just outside the output surface, $\langle I(t)\rangle$, is obtained by taking the Fourier transform of $G(L,z'$=$\ell;\Omega)$ in $\Omega$. A fit of the expression for $\langle I(t)\rangle$ at early times to the measured data for samples with $L$=61 cm, with $\ell$ and $R$ as fitting parameters, gives $\ell$=2.0 cm and $R$=0.64, and is shown in the inset of Fig.~1(b). Excellent agreements between the SCLT and experiment is found for $t$$<$$4\tau_D$. To compare the long-time behavior, we present these results in a semi-log plot in Fig.~1(b). Also shown in Fig.~1(b) is the result of diffusion theory in which localization effects are absent. The surprising agreement between diffusion theory, SCLT, and measurement for $t$$<$$2\tau_D$ suggests that the renormalization of the diffusion constant is negligible at early times even for localized waves studied here. The applicability of diffusion theory at early times in our samples can also be seen in the excellent agreement between the measured and calculated peak arrival times, $t_p$, shown in Fig.~2. We believe, this is because $\tau_D$$<$$\tau_{\rm H}$$<$$\tau_{\rm Th}$ in our samples. The first inequality guarantees that the wave propagation is diffusive for early times, while the second indicates that the samples are in the localized regime. For example, we have $\tau_D$=70.4 ns, $\tau_{\rm H}$=528 ns, and $\tau_{\rm Th}$=695 ns for $L$=61cm and $\tau_D$=136.3 ns, $\tau_{\rm H}$=780 ns, and $\tau_{\rm Th}$=1346 ns for $L$=90cm, respectively. The values of $\tau_{\rm H}$ are obtained from measurements of the average spacing between modes in a closed sample \cite{PRL2001}.
\begin{figure}[t!]
\includegraphics[width=\columnwidth]{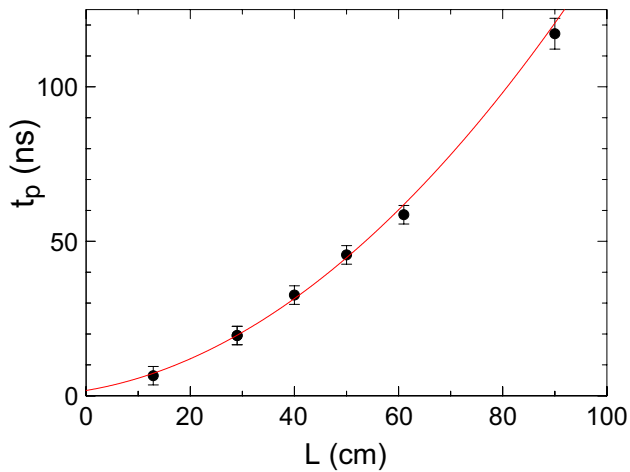}
\caption{(color on line) Measured transmission peak arrival time in alumina samples (solid circles), following incident Gaussian pulse with $\sigma$$\ll$$\tau_D^{-1}$, and the prediction of diffusion theory, $t_p=0.89\tau_D$, with $D$=7.9 cm$^2$/ns and $z_0$=6.1 cm (solid line).}
\end{figure}

Since the SCLT includes localization effects, it gives a slower-than-exponential decay in $\langle I(t)\rangle$ seen in Fig.~1(b). However, for $t$$>$$4\tau_D$, the SCLT underestimates the localization effects and thus gives a higher decay rate than measured. To show this more clearly, we plot both the measured and calculated decay rates in Fig.~3. It is also seen that, at a given delay time, a larger deviation is found for the $L$=61 cm sample than for the $L$=90 cm sample which is deeper in the localization regime. This is because, for a given delay, the ratio of pathlength to sample volume is greater in the shorter sample, so that the number of closed loops and the consequent renormalization is greater in the shorter sample. A slower decay in both samples suggests that more extended quasimodes that may exhibit diffusion-like behavior, and which can be described by the SCLT, have largely decayed so that the wave is transmitted increasingly via long-lived localized modes.
\begin{figure}[t!]
\includegraphics[width=\columnwidth]{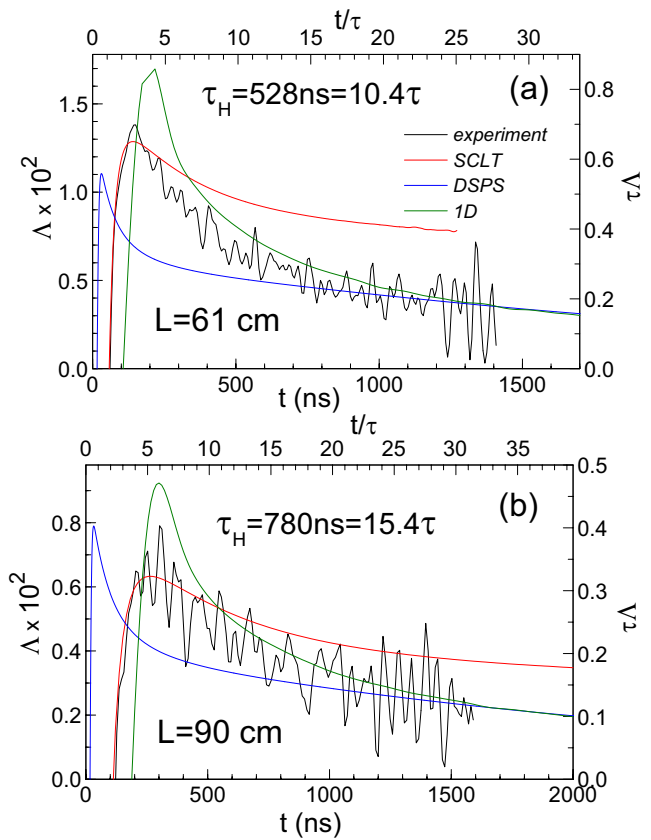}
\caption{(color on line) The transmission decay rate, $\Lambda(t)$, measured in samples with $L$=61 (a) and 90 cm (b) is compared to predictions of the SCLT and dynamic SPS model. The comparison to 1D simulations is made by using the dimensionless decay rate, $\tau\Lambda$, plotted as a function of the dimensionless time, $t/\tau$, where $\tau$=$\bar{\xi}/\beta v_E$=50.6 ns.}
\end{figure}

Because transmission at long times may be determined by the longitudinal structure of spectrally isolated localized modes, we consider the statistics of such modes. In 1D, the steady-state intensity at $z$=$L$ of resonantly excited modes relative to the incident wave at $z$=0$_+$, for modes peaked a distance $z$ from either sample boundary, was given by Azbel \cite{Azbel}, $T$=$\exp(-2\gamma(L-2z))$. The Lyapunov exponent, $\gamma$=1/2$\xi$, is drawn from a Gaussian distribution, $P(\gamma)$=$\sqrt{L/2\pi\bar{\gamma}}\exp[-(\gamma-\bar{\gamma})^2/(2\bar{\gamma}/L)]$, according to the SPS hypothesis \cite{SPS}. We assume the position of peak intensity for the modes is uniformly distributed between 0 and $L$. The decay rate of localized states is the ratio of the sum of the outgoing flux at the open ends to the integrated wave energy inside the sample,
\begin{equation}
\Gamma(\gamma,z)=\beta v_E{1+\exp(-2\gamma(L-2z))\over[2\exp(2z\gamma)-\exp(-2\gamma(L-2z))-1]/2\gamma}\, .
\label{}
\end{equation}
The coupling factor, $\beta$, reflects the reduction of the transmitted flux due to the angular spread about the normal to the interface as well as the angular average of internal reflection at the interface, $R$, and by the character of transport near the boundary. The transmitted intensity is then,
\begin{equation}
\langle I(t)\rangle={1\over 2L}\int_{2\!/L}^{\infty}\!\! d\gamma\!\!\int_0^{L\!/2}\!\!dz P(\gamma)\,T\,\Gamma^2\exp(-\Gamma t)\, .
\label{}
\end{equation}
The factor $T\,(\Gamma/2)^2\exp(-\Gamma t)$ is the square of the Fourier transform of a Lorentzian line in the field spectrum with linewidth $\Gamma/2\pi$. Internal reflectivity will not change $T$ appreciably since the intensity throughout the sample is enhanced by the same factor that inhibits transmission at the interface. By using $\bar{\xi}$=30 cm \cite{PRL2001}, we fit $\Lambda(t)$ calculated from Eq.~(4) to measurements in the samples with $L$=61 and 90 cm, to find $\beta$=0.05. The small coupling factor at the boundary is a consequence both of the strong reflection and the suppressed flow of energy in the exponential tail of localized modes. These results are shown in Fig.~3. The rise time of transmission in this model is the rise time of a localized mode, which is essentially the rise time of the incident pulse. Thus the slower rise observed in the experiment indicates the dominance of overlapping as opposed to isolated modes. At long times, $t$$>$$\tau_{\rm H}$, however, the excellent agreement of the dynamic SPS model with measurements indicates that the energy within the sample is stored in long-lived localized modes.

The above calculations of localized mode dynamics may be compared to 1D simulations in samples with the same values of $L/\bar{\xi}$. We consider a random sample of $L/a$ layers with equal thickness, $a$, embedded in air. The dielectric constant in each layer is a random number uniformly distributed about $\epsilon$=1, from 0.3 to 1.7. A Gaussian pulse with carrier frequency of $\omega_0$=1.65$c/a$, where $c$ is the speed of light, and width $\sigma$=0.14$c/a$ is incident upon the sample, and the intensity $I(t)$ just beyond the output surface is calculated. Over the width of the incident spectrum, $\bar{\xi}$=22$a$. $\langle I(t)\rangle$ is obtained by averaging over 10,000 configurations. The comparison of propagation in quasi-1D and 1D samples is facilitated by employing a dimensionless time $t/\tau$, where $\tau$=$\bar{\xi}/\beta v_E$ [Fig.~3]. The dimensionless decay rate, $\tau\Lambda(t)$, obtained from Eq.~(4) is a universal function of $t/\tau$, which depends only on the ratio $L/\bar{\xi}$. In 1D systems, we take $\beta$=1, since the average index of the sample is the same as its surroundings and no angle average is necessary. Further, we assume $v_E$=$c$, the effective medium velocity, since there are no internal resonances over the pulse bandwidth. The greater delay of the peak in $\tau\Lambda$ and the higher decay rate after the peak found in 1D simulations relative to the quasi-1D measurements, as seen in Fig.~3, indicate that overlapping quasi-extended modes \cite{Pendry,Sebbah,Wiersma} which produce diffusion-like behavior at early times are more prominent in 1D than in quasi-1D.

In conclusion, we find that the nature of pulsed transmission in localized samples evolves with time. At short times, $t$$<$$2\tau_D$, propagation is diffusive; at intermediate times, $2\tau_D$$<$$t$$<$$4\tau_D$, transport can be described in terms of a position and frequency dependent renormalized diffusion coefficient, while at later times,$t$$>$$\tau_{\rm H}$, energy flows exclusively from isolated localized modes. This transition mirrors a change in the distribution of modes over time with the distribution flowing from short-lived overlapping modes towards long-lived localized modes with increasing time. This work shows that a unified theoretical description of pulsed propagation of localized waves will need to incorporate the full distribution of spacings and widths of quasimodes of the random medium as a function of the average overlap parameter, $\delta$$=$$\delta\nu/\Delta\nu$.

We thank Sheng Zhang for sharing results on internal reflection. This research is sponsored by the National Science Foundation under grant number DMR-0538350 and by the Hong Kong RGC under grant number 604506.

\end{document}